\documentclass[aps,prd,notitlepage,amssymb,amsmath,floatfix,nofootinbib,superscriptaddress]{revtex4-1}

\usepackage{epsfig}
\usepackage{bm}
\usepackage{amssymb}
\usepackage{amsmath}
\usepackage{color}
\usepackage{slashed}
\usepackage{subfigure}
\usepackage[colorlinks,linkcolor=blue,anchorcolor=black,citecolor=blue]{hyperref}
\allowdisplaybreaks[4]
\usepackage{comment}
\usepackage{tabularx}
\usepackage{hhline}

\begin{document}

\title{Transverse spin correlation of back-to-back dihadron in unpolarized collisions}

\author{Lei Yang}
\affiliation{Institute of Frontier and Interdisciplinary Science, Key Laboratory of Particle Physics and Particle Irradiation (MOE), Shandong University, Qingdao, Shandong 266237, China}

\author{Yu-Kun Song}
\email{sps\_songyk@ujn.edu.cn}
\affiliation{School of Physics and Technology, University of Jinan, Jinan, Shandong 250022, China}

\author{Shu-Yi Wei}   
\email{shuyi@sdu.edu.cn}
\affiliation{Institute of Frontier and Interdisciplinary Science, Key Laboratory of Particle Physics and Particle Irradiation (MOE), Shandong University, Qingdao, Shandong 266237, China}

\begin{abstract}
The spin correlation of back-to-back dihadron emerges in unpolarized high-energy collisions, empowering unpolarized experiments to shed light on the spin-dependent fragmentation functions. This work investigates the transverse spin correlation of back-to-back dihadron in unpolarized $e^+e^-$, $pp$, and $\gamma p$ collisions, which serves as a novel probe of the chiral-odd fragmentation function $H_{1T}(z)$. We compute the transverse spin correlation at the partonic level and establish a connection with helicity amplitudes. Measuring this observable in future experiments can reveal valuable information on the hadronization of transversely polarized quarks.
\end{abstract}

\maketitle

\section{Introduction}

The hadronization of a high energy parton, described by fragmentation functions in QCD factorization theorem, is a fundamental element in understanding color confinement. The study of fragmentation functions consists of various interesting topics \cite{Metz:2016swz, Chen:2023kqw, Boussarie:2023izj}, and each topic amounts to a tree in the forest. For instance, the unpolarized fragmentation function, $D_1 (z)$, describes the momentum distribution of produced hadron, and the longitudinal spin transfer $G_{1L} (z)$ and the transverse spin transfer $H_{1T} (z)$ deliver complementary information on the transition of spin polarization from the fragmenting parton to the produced hadron. Piecing those parts together eventually could achieve a key milestone in establishing a complete dynamic picture of hadronization mechanism. 

The unpolarized fragmentation function has been extensively investigated in the last decades \cite{Binnewies:1994ju, Kniehl:2000fe, Kneesch:2007ey, Kretzer:2000yf, Albino:2005me, Albino:2005mv, Albino:2008fy, deFlorian:2007aj, deFlorian:2007ekg, Hirai:2007cx, Aidala:2010bn, deFlorian:2014xna, deFlorian:2017lwf, Bertone:2017tyb, Khalek:2021gxf, arXiv:1905.03788, arXiv:2101.04664, arXiv:2210.06078}. As it happens, recent progress \cite{Gao:2024nkz, Gao:2024dbv} even makes precision studies possible. In contrast, the spin-dependent fragmentation functions was mainly investigated in polarized experiments \cite{Ma:1999wp, Ma:2000uu, Ma:2001na}, such as LEP and RHIC. Very few colliders in the world are capable of conducting such research. Therefore, model calculations \cite{Nzar:1995wb, Metz:2002iz, Bacchetta:2007wc, Gamberg:2008yt, Lu:2015wja, Yang:2016mxl, Yang:2017cwi} have played an important role in theoretical studies. The perspective of understanding hadronization from the spin degree of freedom has been largely arrested mainly due to the difficulties in experimental measurements. On top of this, Ref.~\cite{Pan:2023pve} also demonstrated that the decay contribution often contaminates the quantitative study of spin-dependent fragmentation functions, and therefore makes precision measurements even more challenging.

A pioneering study \cite{Chen:1994ar} proposed to measure the longitudinal spin transfer $G_{1L} (z)$ at unpolarized electron-positron colliders utilizing the helicity correlation of back-to-back dihadron. Established on a cascade of recent studies \cite{Zhang:2023ugf, Li:2023qgj, Chen:2024qvx}, it comes to light that helicity correlation is a direct consequence of partonic hard interaction and exhibits in all high energy collisions. Therefore, they also explored the opportunity of investigating the longitudinal spin transfer in unpolarized $pp$, $AA$, and $ep$ collisions. In light of the amazing progress of experimental measurements on dihadron spin-spin correlation \cite{Gong:2021bcp, Vanek:2023oeo}, this novel observable offers hope to study spin-dependent fragmentation functions in unpolarized colliders. A global analysis in the future could significantly broaden our knowledge of the hadronization mechanism. Moreover, the spin correlation has recently inspired more proposals \cite{Tu:2023few, Barata:2023jgd, Shao:2023bga, Lv:2024uev, Wu:2024mtj, Wu:2024asu, Shen:2024buh} in various contexts and emerges as a new frontier whose potential still has not been fully unleashed yet.

Besides the helicity correlation, the back-to-back partons also evince transverse spin correlation \cite{Chen:1994ar}, which translates into that of dihadron through hadronization. The hadronization of a transversely polarized quark is described by chiral-odd fragmentation functions. {In the TMD factorization, it usually leads to azimuthal asymmetries thanks to the renowned Collins function. Refs.~\cite{Collins:1993kq, Collins:1994ax, Jaffe:1997hf, Bianconi:1999cd, Radici:2001na, Bacchetta:2002ux, Boer:2003ya, Bacchetta:2008wb, Matevosyan:2018icf} also established a connection between transversely polarized partons with dihadron fragmentation function where two partons are produced by the same parton, while Refs.~\cite{Bacchetta:2004it, Courtoy:2012ry, Radici:2018iag} extracted those dihadron fragmentation functions from experimental data. Recent work \cite{Wen:2024cfu} employed this quantity to investigate beyond standard model theories. Notice that the word ``{\it dihadron}'' has multiple meanings. In the context of dihadron fragmentation function, it refers two neighboring hadrons residing in the same jet-cone, since they are produced by the same parton. The back-to-back dihadron in our paper refers to two hadrons produced by different partons. Therefore, in the transverse plane, they are almost back-to-back.} In the collinear factorization of single hadron fragmentation functions, only the transverse spin transfer $H_{1T}$ contributes, akin to the transversity $h_{1T} (x)$ of parton distribution function \cite{Ralston:1979ys, Artru:1989zv, Jaffe:1991kp, Jaffe:1991ra, Cortes:1991ja}. However, it is important to note that the helicity and transverse spin correlations are two distinctly different quantities which are only loosely related to each other by the positivity constraint. They are {\it not} different projections of the same quantity, and therefore we cannot derive one from the other. The reasons are listed in the following.

First, the helicity correlation of two partons exists as long as they are interacting with each other. On the contrary, as demonstrated later, the transverse spin correlation of two interacting partons only manifests in a selection of partonic channels. The unconnected diagrams cannot contribute to transverse spin correlation. In the language of helicity amplitude approach, the helicity correlation measures the difference between the magnitudes of two amplitudes, while the transverse spin correlation quantifies their interference. 

Furthermore, it is also not feasible to establish a naive connection between the transverse spin transfer $H_{1T}$ and its longitudinal counterpart $G_{1L}$. For instance, both quark and gluon contribute to the longitudinal spin transfer. (The partner of the longitudinally polarized quark is the {\it circularly polarized gluon}.) However, for the transverse spin transfer, only the quark sector contributes. The linearly polarized gluon, which is the counterpart of the transversely polarized quark, cannot contribute to the transverse polarization of produced hadrons \cite{Jaffe:1989xy, Ji:1992eu, Soffer:1997zy, Boussarie:2023izj} in the collinear factorization. As a result, the DGLAP evolution \cite{Altarelli:1977zs} of $H_{1T}$ becomes a diagonal one \cite{Vogelsang:1997ak, Barone:2001sp}. The gluon sector vanishes. This feature grants the $H_{1T}$ fragmentation function a unique advantage in understanding the hadronization mechanism: the perfect separation between quark and gluon contributions. Therefore, the aim of this paper is to investigate the transverse spin correlation in unpolarized high energy collisions and to improve our quantitative understanding of the chiral-odd $H_{1T}$ fragmentation function.

The rest of this paper is organized as follows. In Sec. II, we first study the most simple case: the transverse spin correlation in unpolarized electron positron annihilation process. In Sec. III, we present the transverse spin correlation in unpolarized hadronic collisions. In Sec. IV, we present that in photon-nucleus collisions, which can be directly applied to the ultra-peripheral nucleus-nucleus collision or the electron-ion collisions. We present the relation between the transverse spin correlation and helicity amplitudes in Sec. V and give a summary in Sec. VI.

\section{Transverse spin correlation in $e^+ e^-$ annihilation}
\label{sec:ee}

We first consider the simple back-to-back $\Lambda$-$\bar\Lambda$ pair production in $e^+e^-$ annihilations, which was first calculated in Ref.~\cite{Chen:1994ar}, to demonstrate the origin of the transverse spin correlation in this section, and then extend this approach to the unpolarized pp collisions in the next section. 

\begin{figure}[htb]
\includegraphics[width=0.4\textwidth]{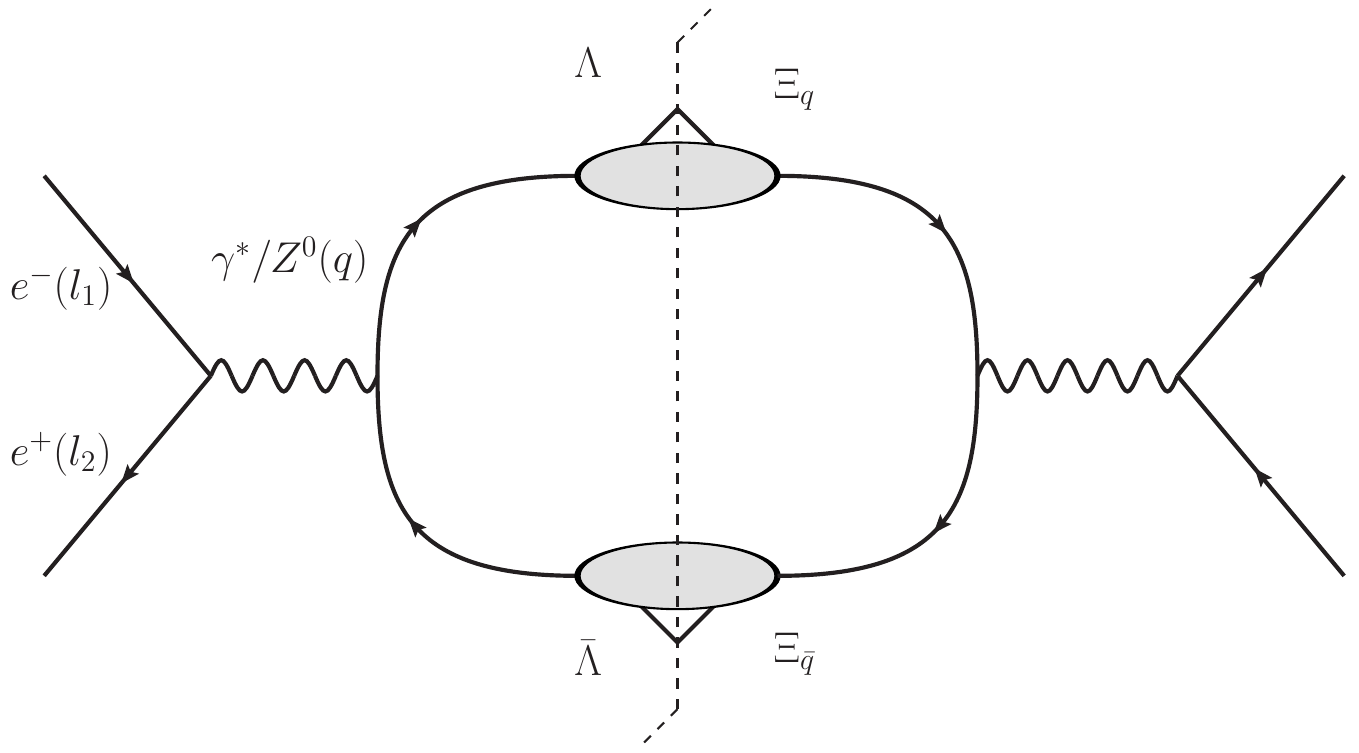}
\caption{The leading order Feynman diagram for the $\Lambda$-$\bar\Lambda$ pair production in $e^+e^-$ annihilation.}
\label{fig:ee}
\end{figure}

As illustrated in Fig.~\ref{fig:ee}, the LO contribution comes from $e^-(l_1) + e^+ (l_2) \to q (\to \Lambda) + \bar q(\to \bar\Lambda)$. In principle, we also need to consider the contribution from $q(\bar\Lambda) + \bar q(\to \Lambda)$ as well, which is rather straightforward.  We will evaluate this part of the contribution when presenting the final result. 

In $e^+ e^-$ annihilation, we can first reconstruct the thrust axis which represents the quark and antiquark direction at LO. The beam axis $\bm{n}_{e^+e^-}$ and the thrust axis $\bm{n}_\text{thrust}$ thus span the production plane. The transverse spin correlation of the final state dihadron is investigated along the normal direction of this hadron production plane, i.e., $\bm{n}_T = {\bm{n}}_{e^+e^-} \times {\bm{n}}_\text{thrust}$. Due to the parton shower, the final state $\Lambda$ hyperons do not move exactly along the thrust axis. They differ with each other by a very small angle at high energy limit. Although it is more convenient to measure the transverse spin correlation along the transverse direction defined by $\bm{n}_T' = \hat{\bm{p}}_\Lambda \times \hat{\bm{p}}_{\bar\Lambda}$ with $\hat{\bm{p}}_{\Lambda/\bar\Lambda}$ the unit vector along the $\Lambda/\bar\Lambda$ momentum direction respectively, it might receive contribution from the $D_{1T}^\perp$ fragmentation function which is responsible for producing transversely polarized hadron from unpolarized parton. Similar to the helicity correlation defined in Refs.~\cite{Zhang:2023ugf, Li:2023qgj, Chen:2024qvx}, the transverse spin correlation is defined as 
\begin{align}
{\cal C}_{TT} = \frac{{\cal P} (\bm{n}_T , \bm{n}_T ) + {\cal P} (-\bm{n}_T , -\bm{n}_T ) - {\cal P} (\bm{n}_T , -\bm{n}_T ) - {\cal P} (-\bm{n}_T , \bm{n}_T )}{{\cal P} (\bm{n}_T , \bm{n}_T ) + {\cal P} (-\bm{n}_T , -\bm{n}_T ) + {\cal P} (\bm{n}_T , -\bm{n}_T ) + {\cal P} (-\bm{n}_T , \bm{n}_T )},
\end{align}
where ${\cal P} (\bm{a}_T , \bm{b}_T )$ is the probability for $\Lambda$ being transversely polarized along the $\bm{a}_T $ direction and $\bar \Lambda$ being transversely polarized along the $\bm{b}_T$ direction. 

In the parton model, the LO differential cross section reads
\begin{align}
& \frac{d\sigma}{dy dz_1 dz_2 d^2\bm{P}_\perp} = \frac{2\pi N_c \alpha^2}{Q^4} L_{\mu\nu} W^{\mu\nu},
\end{align}
where $Q$ is the center-of-mass energy of $e^+e^-$, $y=(1+\cos\theta)/2$ with $\theta$ the angle between $e^-$ and $\Lambda$, $z_{1,2}$ is the light-cone momentum fraction of the final state $\Lambda/\bar\Lambda$, and $\bm{P}_\perp$ is the relative transverse momentum of $\Lambda$ with respect to the $\bar\Lambda$ momentum. Here, we have only consider the electromagnetic interaction for simplicity. The weak interaction will be taken into account eventually when presenting our final result. The leptonic tensor $L_{\mu\nu}$ and the hadronic tensor $W^{\mu\nu}$ are given by
\begin{align}
&
L_{\mu\nu} = l_{1\mu}l_{2\nu} + l_{1\nu} l_{2\mu} - g_{\mu\nu} l_1 \cdot l_2,
\\
&
W^{\mu\nu} = \sum_q e_q^2 \int d^2 \bm{p}_{T1} d^2 \bm{p}_{T2} \delta^2 (\bm{P}_\perp - \bm{p}_{T2} - \frac{z_2}{z_1} \bm{p}_{T1}) {\rm Tr} \left[ 2 \Xi_q^{\Lambda} (z_1, \bm{p}_{T1}) \gamma^\mu 2 \Xi_{\bar q}^{\bar\Lambda} (z_2, \bm{p}_{T2}) \gamma^\nu \right],
\end{align}
with $e_q$ being the electric change of quark $q$. Here $\Xi_{q}^{\Lambda}$ and $\Xi_{\bar q}^{\bar\Lambda}$ are four by four matrices which can further be decomposed in terms of Gamma matrices. The leading twist decomposition for spin-$1/2$ baryon production contains eight structures which can be found in Refs.~\cite{Metz:2016swz, Chen:2023kqw, Boussarie:2023izj, Wei:2013csa, Wei:2014pma}. We have omitted the other terms since they disappear after integrating over the relative transverse momentum, and focus only on the transverse spin correlation. Therefore, we only keep the following three structures
\begin{align}
& \Xi_{q}^{\Lambda} (z_1, \bm{p}_{T1}) = \frac{1}{4} \slashed n_+ D_{1,q}^{\Lambda} (z_1, \bm{p}_{T1}) + \frac{1}{8} [\slashed S_{T1}, \slashed n_+] \gamma_5 H_{1T,q}^{\Lambda} (z_1, \bm{p}_{T1}) + \frac{{p}_{T1} \cdot {S}_{T1}}{8M_1^2} [\slashed p_{T1}, \slashed n_+] \gamma_5 H_{1T,q}^{\perp,\Lambda} (z_1, \bm{p}_{T1}) , 
\\
& \Xi_{\bar q}^{\bar\Lambda} (z_2, \bm{p}_{T2}) = \frac{1}{4} \slashed n_- D_{1,\bar q}^{\bar\Lambda} (z_2, \bm{p}_{T2}) + \frac{1}{8} [\slashed S_{T2}, \slashed n_-] \gamma_5 H_{1T, \bar q}^{\bar\Lambda} (z_2, \bm{p}_{T2}) + \frac{{p}_{T2} \cdot {S}_{T2}}{8M_2^2} [\slashed p_{T2}, \slashed n_-] \gamma_5 H_{1T,\bar q}^{\perp,\bar\Lambda} (z_2, \bm{p}_{T2}) ,
\end{align}
where $D_{1}$ is the unpolarized TMD fragmentation function and $H_{1T}$ and $H_{1T}^\perp$ are chiral-odd TMD fragmentation functions describing the hadronization of transversely polarized partons. Notice that this decomposition is convention-dependent. In this work, we follow the convention from Refs.~\cite{Metz:2016swz, Chen:2023kqw}. However, one can always allocate some contributions from $H_{1T}^\perp$ to $H_{1T}$. For instance, Ref.~\cite{Boussarie:2023izj} offers a rather complicated Lorentz structure in the TMD factorization. However, as shown later, our {\it simple} TMD structure~\cite{Metz:2016swz, Chen:2023kqw} leads to an {\it involved} relation between TMD fragmentation functions and the collinear ones. On the contrary, the {\it involved} TMD structure in Ref.~\cite{Boussarie:2023izj} leads to a {\it simple} relation. The physical interpretations of chiral-odd fragmentation functions in different conventions are also different. They also follow different evolution equations. Nonetheless, the final results of observables remain convention-independent, since we have simply shuffled some contribution from one part to another. The sum remains the same. 

Furthermore, integrating over the relative transverse momentum between two hadrons $\bm{P}_\perp$, we recover the expression in collinear factorization. The cross section is then given by
\begin{align}
& \frac{d\sigma}{dy dz_1 dz_2} = \frac{2\pi N_c \alpha^2}{Q^2} L_{\mu\nu} \hat W^{\mu\nu}, \label{eq:collinear-ee}
\end{align}
with the $\bm{P}_\perp$-integrated hadronic tensor being given by
\begin{align}
&
\hat W^{\mu\nu} = \sum_q e_q^2 {\rm Tr} \left[ 2 \hat\Xi_q^{\Lambda} (z_1) \gamma^\mu 2 \hat \Xi_{\bar q}^{\bar\Lambda} (z_2) \gamma^\nu \right].
\end{align}
Here, $\hat \Xi_{q,\bar q} (z)$ is the $\bm{p}_T$-integrated version of $\Xi_{q,\bar q}(z,\bm{p}_T)$. The decomposition in the collinear factorization reads
\begin{align}
& \hat \Xi_{q}^{\Lambda} (z_1) = \frac{1}{4} \slashed n_+ D_{1,q}^{\Lambda} (z_1) + \frac{1}{8} [\slashed S_{T1}, \slashed n_+] \gamma_5 H_{1T,q}^{\Lambda} (z_1) , 
\\
& \hat \Xi_{\bar q}^{\bar\Lambda} (z_2) = \frac{1}{4} \slashed n_- D_{1,\bar q}^{\bar\Lambda} (z_2) + \frac{1}{8} [\slashed S_{T2}, \slashed n_-] \gamma_5 H_{1T, \bar q}^{\bar\Lambda} (z_2),
\end{align}
with $D_1$, $H_{1T}$ being collinear fragmentation functions which can be related to TMD ones by
\begin{align}
& D_1 (z) = \int d^2 \bm{p}_T D_1 (z, \bm{p}_T),
\\
& H_{1T} (z) = \int d^2 \bm{p}_T \left[ H_{1T} (z, \bm{p}_T) + \frac{{p}_{T}^2}{2M^2} H_{1T}^\perp (z,\bm{p}_T) \right].
\end{align}
While $D_1(z)$ is the unpolarized cross section, $H_{1T} (z)$ is the transverse spin transfer representing the number density of producing transversely polarized hadron from transversely polarized quark. As discussed above, the contribution from $H_{1T}^\perp$ term does not vanish in our TMD convention. This leads to the strange relation between the collinear function $H_{1T}(z)$ with the TMD functions $H_{1T}(z,\bm{p}_T)$ and $H_{1T}^\perp(z,\bm{p}_T)$. Nonetheless, if we adopt the convention from Ref.~\cite{Boussarie:2023izj}, we will find that the contribution from $H_{1T}^\perp$ disappears integrating over the transverse momentum. Thus, we can establish a simple relation between $H_{1T}(z)$ and $H_{1T}(z,\bm{p}_T)$. 

Inserting the leptonic tensor and the hadronic tensor into Eq.~(\ref{eq:collinear-ee}) and taking into account both electromagnetic and weak interactions, we obtain
\begin{align}
\frac{d\sigma}{dydz_1 dz_2} = \frac{2\pi N_c \alpha^2}{Q^2} & \Bigg\{ \sum_q \left[ 
  \omega_q (y) D_{1,q}^{\Lambda} (z_1) D_{1,\bar q}^{\bar\Lambda} (z_2) 
+ (\bm{S}_{T1} \cdot \bm{S}_{T2}) \omega_q^T (y)  H_{1T,q}^{\Lambda} (z_1) H_{1T,\bar q}^{\bar\Lambda} (z_2) 
\right] \nonumber\\
+ & \sum_q \left[ 
  \omega_q (1-y) D_{1,\bar q}^{\Lambda} (z_1) D_{1,q}^{\bar\Lambda} (z_2) 
+ (\bm{S}_{T1} \cdot \bm{S}_{T2}) \omega_q^T (1-y)  H_{1T,\bar q}^{\Lambda} (z_1) H_{1T,q}^{\bar\Lambda} (z_2) 
\right] \Bigg\}
,\label{eq:final-ee}
\end{align}
where the coefficient functions are consistent with those in Refs.~\cite{Wei:2013csa, Wei:2014pma, Chen:2016moq, Chen:2021zrr, Boer:1997mf} and also are listed in Appendix \ref{sec:coeff} for self-sufficient. This result is also consistent with that in Ref.~\cite{Chen:1994ar}. Both $D_{1}$ and $H_{1T}$ are scale dependent, which can be obtained by solving the DGLAP evolution equation. However, the polarized splitting functions differ from the unpolarized ones. We explicitly lay out the evolution equation of $H_{1T}(z,\mu^2)$ and investigate the impact on the transverse spin transfer in Appendix \ref{sec:evolution}. Moreover, we have used the product of three-vectors in the above expression, $\bm{S}_{T1} \cdot \bm{S}_{T2}$. Since we are only interested in the correlation of transverse polarization along the normal direction of the hadron production plane, $\bm{S}_{T1} \cdot \bm{S}_{T2}$ could either be $+1$ for same sign polarizations or $-1$ for opposite sign polarizations. Notice that if $\bm{n}_T$ is not perpendicular to the lepton momentum, other terms may affect the transverse spin correlation as well. Moreover, the second line computes the contribution from the less important $q \to \bar\Lambda, \bar q\to \Lambda$ channel. At the end of the day, the dihadron transverse spin correlation ${\cal C}_{TT}$ is evaluated by
\begin{align}
{\cal C}_{TT} (y, z_1, z_2) = \frac{\sum_q \left[ \omega_q^T (y) H_{1T,q}^{\Lambda} (z_1) H_{1T,\bar q}^{\bar\Lambda} (z_2) + \omega_q^{T} (1-y) H_{1T,\bar q}^{\Lambda} (z_1) H_{1T, q}^{\bar\Lambda} (z_2) \right] }{\sum_q \left[ \omega_q (y) D_{1,q}^{\Lambda} (z_1) D_{1,\bar q}^{\bar\Lambda} (z_2) + \omega_q (1-y) D_{1,\bar q}^{\Lambda} (z_1) D_{1,q}^{\bar\Lambda} (z_2) \right]}.
\end{align}
Here, we have already taken into account contribution from $(q \to \bar\Lambda, \bar q\to \Lambda)$. The physical interpretation of this result is also clear. $\omega_q^T/\omega_q$ is the transverse spin correlation of $q\bar q$ pair. The partonic correlation can be inherited by final state hadrons by convoluting the numerator and the denominator with the corresponding fragmentation functions. By measuring the transverse spin correlation at the Belle/LEP experiments, we can investigate the chiral-odd $H_{1T} (z)$ fragmentation function. 

\begin{figure}[h!]\centering
\includegraphics[width=0.45\textwidth]{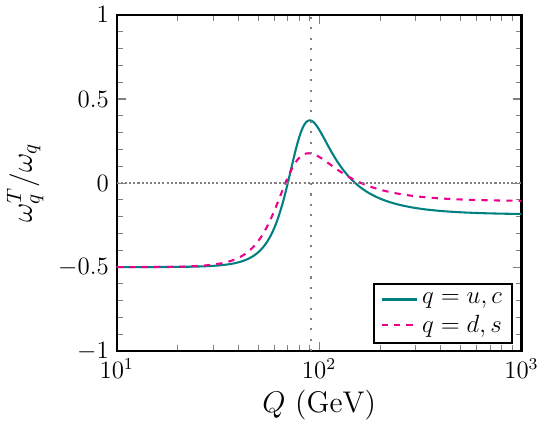}
\caption{Transverse spin correlation of final state $q\bar q$ pair in unpolarized $e^+e^-$ collisions as a function of $Q$.}
\label{fig:spin-cor-ee}
\end{figure}

Since our quantitative knowledge on $H_{1T} (z)$ is next to nothing, we only present our numerical results at the partonic level. The transverse spin correlation of the final state $q\bar q$ pair in unpolarized electron-positron annihilation experiments is shown in Fig.~\ref{fig:spin-cor-ee}. In the numerical evaluation, we have integrated over $y$ in the allowed kinematic region. There is a sign flip between low-energy experiments, such as Belle, and high energy experiments, such as LEP. The sign flip stems from the competition between the electromagnetic $\gamma^\mu$ vertex and the weak $c_V \gamma_\mu - c_A \gamma_\mu\gamma_5$ vertex. 

Thanks to the positivity constrain $|{\cal C}_{TT}| \le \sqrt{1 - |{\cal P}_L|^2}$ with $|{\cal P}_L|$ the magnitude of the helicity of quark/antiquark, the transverse spin correlation at the $Z^0$ pole is a bit smaller than that at lower energy. Furthermore, since the down-type-quark helicity is larger than that of the up-type quark at the $Z^0$ pole \cite{Augustin:1978wf, Gustafson:1992iq, Chen:2016iey}, the transverse spin correlation of the down-type quark-antiquark pair is smaller than that of the up-type pair. Furthermore, the spin-dependent fragmentation functions also satisfy a positivity constrain as presented in Refs.~\cite{Soffer:1994ww, Vogelsang:1997ak}.

It is also intriguing to note that if the quark-$Z^0$-boson vertex is assumed to be $\gamma_\mu (1 \pm \gamma_5)$, the transverse spin correlation between quark and antiquark reduces to zero. The reason for this phenomenon can be better appreciated in the language of the helicity amplitude approach.  The transverse spin correlation requires a helicity flip between the amplitude and the conjugate amplitude \cite{Hoodbhoy:1998vm}. However, this cannot happen if the coupling vertex takes the $\gamma_\mu (1 \pm \gamma_5)$ form. Furthermore, the vanishing transverse spin correlation can happen in other circumstances as well, which will be elaborated in Sec.~\ref{sec:helicity_amplitude}.

\section{Transverse spin correlation in unpolarized $pp$ collisions}

In this section, we extend our research of the transverse spin correlation to unpolarized $pp$ collisions and explore the opportunity of studying the transverse spin transfer in unpolarized hadron collider experiments such as LHC, RHIC, and Tevatron. Similar to the case in $e^+e^-$ annihilation, the momenta of two final state jets which are almost back-to-back in the transverse plane in $pp$ collisions also span a production plane. We thus can investigate the transverse spin correlation of $\Lambda$-$\bar\Lambda$ pair along the normal direction of the production plane. 

We would like to discuss the difference between the helicity correlation and the transverse spin correlation in $pp$ collisions. All partonic channels contribute to the helicity correlation, albeit the sign varies with channels. To be more specific, the helicities of final state partons take precisely the opposite sign for the $q_i \bar q_i \to q_j \bar q_j$, $gg\to q_i \bar q_i$ and $q_i \bar q_i \to gg$ channels. Therefore, they contribute to the negative helicity correlation, while the other channels prefer the same sign correlations. The partial cancellation among different channels results in a tiny helicity correlation at the hadronic level. Per contra, the transverse spin correlation arises from the chiral-odd $H_{1T}$ fragmentation function. Only connected channels (i.e., the final state quark and/or antiquark are connected by the same trace line) contribute, while the others amount to the total production rate.

Notice that, in principle, the linear polarizations of gluons produced from the $q_i \bar q_i \to gg$ channel are also correlated. Vis-\`a-vis the transverse spin correlation of the quark-antiquark pair, the linear polarization correlation of gluons indicates that the probabilities for the parallel and perpendicular polarizations are different. However, the collinear $H_{1T}$ fragmentation function of gluon does not exist. They can contribute to other observables, such as the tensor polarization of vector mesons \cite{Boer:2016xqr, Boer:2017xpy, Kumano:2019igu, Kumano:2020gfk, Kumano:2020ijt, Kumano:2021fem, Song:2023ooi} etc. In this paper, we only investigate the transverse spin correlation of back-to-back diquark and leave that of gluons for future work. 

After integrating over the relative transverse momenta, we arrive at the cross section in the collinear factorization framework given by
\begin{align}
\frac{d\sigma_{pp\to \Lambda \bar\Lambda +X}}{dy_1 d^2\bm{p}_{T1} dy_2 d^2\bm{p}_{T2}} 
&
= \int \frac{dz_1}{z_1^2} \frac{dz_2}{z_2^2} \sum_{ab\to cd} x_a f_{1,a} (x_a) x_b f_{2,b} (x_b) \delta^2 \left(\frac{\bm{p}_{T1}}{z_1} + \frac{\bm{p}_{T2}}{z_2} \right) 
\nonumber\\
& \times \frac{1}{\pi} \left[
\frac{d\hat\sigma_{ab\to cd}}{dt} D_{1,c}^{\Lambda} (z_1) D_{1,d}^{\bar\Lambda} (z_2) 
+ 
(\bm{S}_{T1} \cdot \bm{S}_{T2}) \frac{d\hat\sigma_{ab\to cd}^T}{dt} H_{1T,c}^{\Lambda} (z_1) H_{1T,d}^{\bar\Lambda} (z_2) 
\right].
\end{align}
Here, $d\hat \sigma_{ab\to cd}/dt$ is the unpolarized cross section of $ab\to cd$ scattering which can be found in Refs.~\cite{Owens:1986mp}, and $d\hat \sigma_{ab\to cd}^T/dt$ is the transversely polarized cross section. The exchange between $(c\to \Lambda, d\to \bar\Lambda)$ and $(c\to \bar\Lambda, d\to \Lambda)$ is implicit.  

First, it is straight forward to find that the $q_i \bar q_i \to q_j \bar q_j$, $gg\to q_i \bar q_i$, and $q_i \bar q_i \to q_i \bar q_i$ channels satisfy the aforementioned criteria. While the unpolarized cross sections of these channels are well known, the transversely polarized cross sections read
\begin{align}
&
\frac{d\hat \sigma_{q_i \bar q_i \to q_j \bar q_j}^T}{dt} = - \frac{2\pi \alpha_s^2}{9s^2} \frac{4ut}{s^2},
\\
&
\frac{d\hat \sigma_{q_i \bar q_i \to q_i \bar q_i}^T}{dt} = \frac{2\pi \alpha_s^2}{9s^2} \frac{4u(s-3t)}{3s^2},
\\
&
\frac{d\hat \sigma_{gg \to q_i \bar q_i}^T}{dt} = \frac{\pi \alpha_s^2}{12s^2} \frac{ut-4u^2-4t^2}{s^2},
\end{align}
with $s,t,u$ Maldanstan variables.

\begin{figure}[h!]
\includegraphics[width=0.45\textwidth]{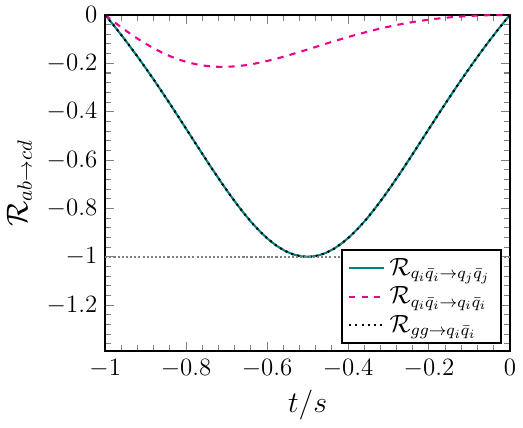}
\caption{The partonic transverse spin correlation $\cal R$ as a function of $t/s$ for $q\bar q$ production channels in $pp$ collisions.}
\label{fig:ratio}
\end{figure}

All three channels are negative indicating that the final state partons prefer to be be transversely polarized along opposite directions. We define ${\cal R}_{ab\to cd}$ as the ratio between the polarized cross section and the unpolarized one, i.e., ${\cal R}_{ab\to cd} \equiv d\hat\sigma_{ab\to cd}^T/d\hat\sigma_{ab\to cd}$, which quantifies the partonic transverse spin correlation. The numerical results are shown in in Fig.~\ref{fig:ratio} as a function of $t/s$. The transverse spin correlation of the $q_i \bar q_i \to q_i \bar q_i$ channel is much smaller than those of $q_i \bar q_i/gg \to q_j \bar q_j$ channels which even reach unity at $t=-s/2$. This is because that the $q_i \bar q_i/gg \to q_j \bar q_j$ channel contains only connected diagrams where final state partons are connected by the same Fermion line. On the other hand, the $q_i \bar q_i \to q_i \bar q_i$ channel contains both $s$ channel and $t$ channel contributions. The unconnected $t$ channel diagram does not contribute to the transverse spin correlation, and therefore reduces the magnitude of the correlation. 

\begin{figure}[h!]
\includegraphics[width=0.3\textwidth]{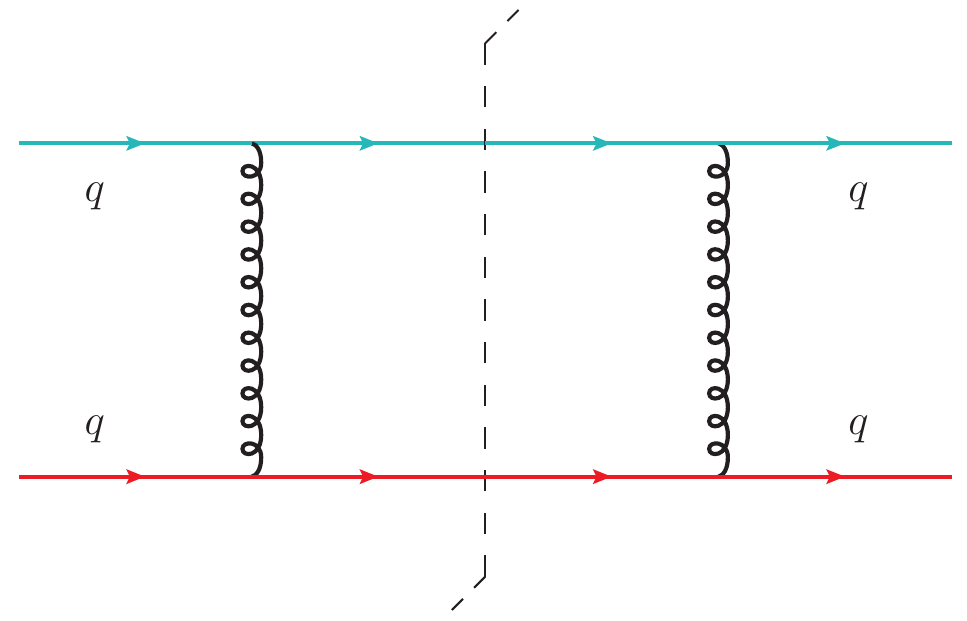}
\phantom{XX}
\includegraphics[width=0.3\textwidth]{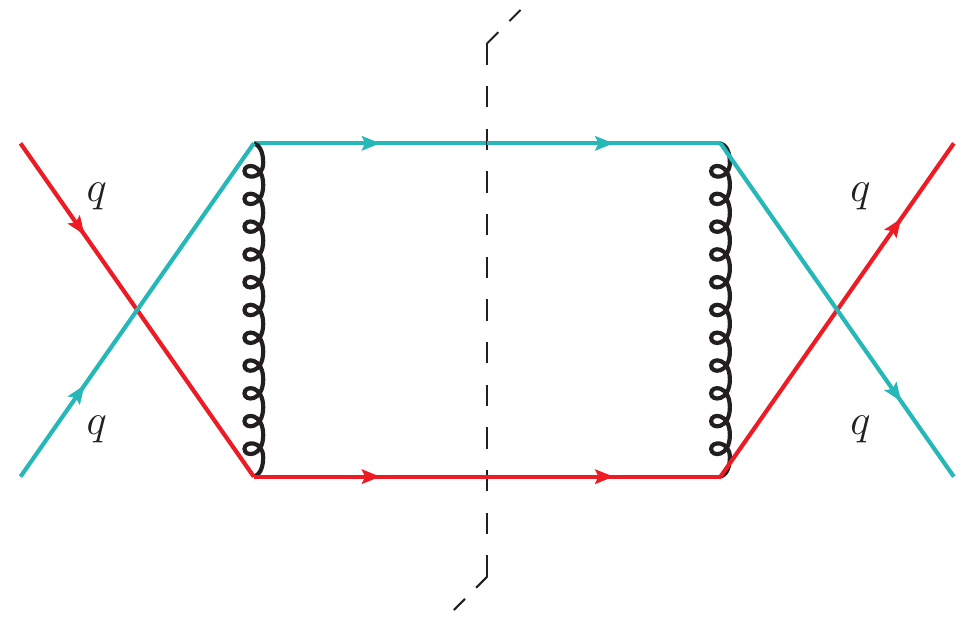}
\phantom{XX}
\includegraphics[width=0.3\textwidth]{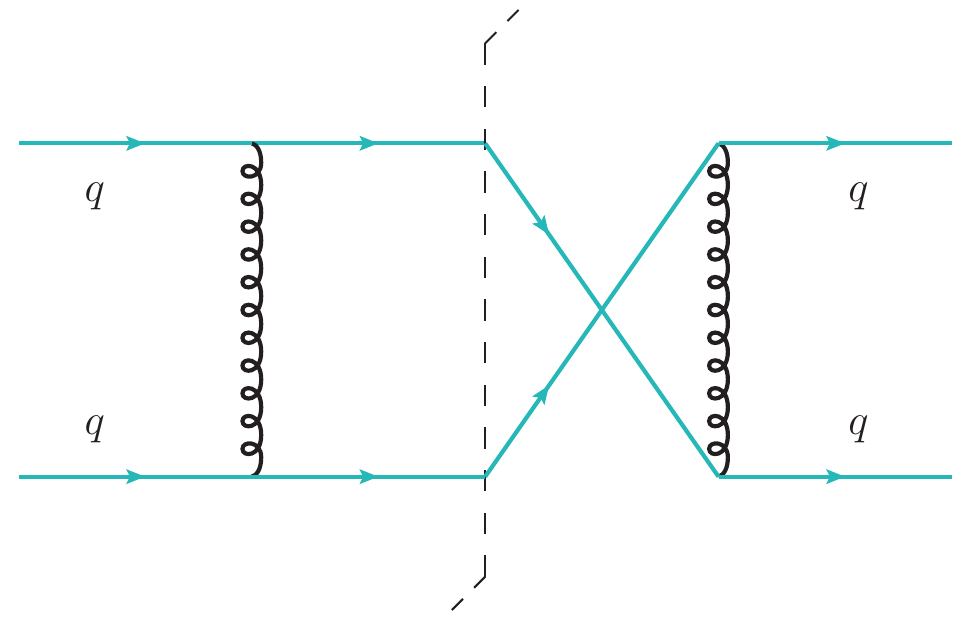}
\caption{Leading order Feynman diagrams for the $qq\to qq$ scatting. Left: $t$-channel diagram; Middle: $u$-channel diagram; Right: the interference diagram. The color of the quark line represents a closed trace line. While the left and middle diagrams do not contribute to the transverse spin correlation, the right one is a connected diagram that contributes.}
\label{fig:pptc}
\end{figure}

Furthermore, the $2\to 2$ process with identical quarks, i.e., $q_i q_i \to q_i q_i$, also contributes. The correlation arises from the interference between $t$-channel and $u$ channel scatterings. As illustrated in Fig.~\ref{fig:pptc}, the left and the middle plots represent the $t$ and $u$ channel diagrams which do not contribute to the transverse spin correlation. The right plot represents the interference between $u$ and $t$ channel scatterings. We have utilized different colors to represent different trace lines. The final state identical quarks in the interference diagram are connected by the same Fermion line. Therefore, it contributes to the transverse spin correlation. We obtain the partonic transverse spin correlation of this channel as
\begin{align}
\frac{d\hat \sigma_{q_i q_i \to q_i q_i}^T}{dt} = - \frac{2\pi \alpha_s^2}{9s^2} \frac{4}{3}.
\end{align}

\begin{figure}[htb]
\includegraphics[width=0.45\textwidth]{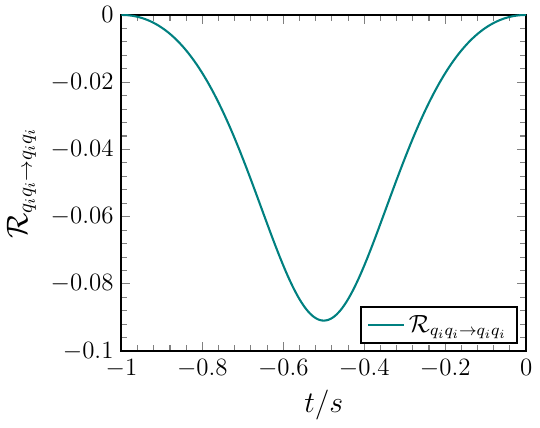}
\caption{The partonic transverse spin correlation $\cal R$ as a function of $t/s$ for the identical-parton-scattering channel in $pp$ collision.}
\label{fig:Rqq}
\end{figure}

We show the numerical result of ${\cal R}_{q_i q_i \to q_i q_i}$ as a function in Fig.~\ref{fig:Rqq}. The magnitude becomes much smaller compared with the other unconnected channels, since it only arises from the interference diagrams. Nonetheless, this channel can become important at the threshold regime with $x_{a,b} \to 1$. 

The transverse spin correlation of final state hadrons is given by convoluting $\sigma^T$ with the transverse spin transfers. We obtain
\begin{align}
{\cal C}_{TT} = \frac{
\int d{\cal P.S.} \sum_{ab\to cd} x_a f_{1,a} (x_a) x_b f_{1,b} (x_b) \frac{1}{\pi} \frac{d\hat \sigma_{ab\to cd}^T}{dt} H_{1T,c}^{\Lambda} (z_1) H_{1T,d}^{\bar\Lambda} (z_2)
}{
\int d{\cal P.S.} \sum_{ab\to cd} x_a f_{1,a} (x_a) x_b f_{1,b} (x_b) \frac{1}{\pi} \frac{d\hat \sigma_{ab\to cd}}{dt} D_{1,c}^{\Lambda} (z_1) D_{1,d}^{\bar\Lambda} (z_2)
},
\end{align}
where $d{\cal P.S.}$ represents the phase space to be integrated over and $f_1 (x_{a,b})$ is the collinear parton distribution function with $x_{a,b}$ the momentum fraction. Since the vast majority of partonic channels do not contribute to the transverse spin correlation, the experimental signal is expected to be small. 

\section{Transverse spin correlation in photon-nucleus collisions}

The high energy large nucleus is accompanied by enormous coherent quasireal photons. In the ultra-peripheral relativistic nucleus-nucleus collisions (UPC), we can have photon-nucleus collisions as discussed in Ref.~\cite{Chen:2024qvx}. At the leading order in QCD, the partonic hard scattering consists of $\gamma g \to q\bar q$ and $\gamma q \to q g$ channels. As discussed in the previous section, the final state gluon does not contribute to the transverse spin transfer. Therefore, only the $\gamma g$ channel contributes to the transverse spin correlation. The $\gamma q$ channel only contributes to the denominator. In the end, the transverse spin correlation in UPC is given by
\begin{align}
{C}_{TT} = \frac{\int d{\cal P.S.} \sum_q x_\gamma f_\gamma (x_\gamma) x_g f_{1,g} (x_g) \frac{1}{\pi}\frac{d\hat \sigma_{\gamma g \to q\bar q}^T}{dt} H_{1T,q}^{\Lambda} (z_1) H_{1T,\bar q}^{\bar\Lambda} (z_2)}{\int d{\cal P.S.} \sum_{b,c,d} x_\gamma f_\gamma (x_\gamma) x_b f_{1,b}(x_b) \frac{1}{\pi}\frac{d\hat\sigma_{\gamma b \to cd}}{dt} D_{1,c}^{\Lambda} (z_1) D_{1,d}^{\bar\Lambda} (z_2)},
\end{align}
where $x_\gamma$ is the per-nucleon momentum fraction carried by the quasireal photon and $f_{\gamma} (x_\gamma)$ is the collinear photon distribution \cite{Jackson:1998nia} which can be expressed as
\begin{align}
x_\gamma f_\gamma (x_\gamma) = \frac{2Z^2 \alpha}{\pi} \left[ \zeta K_0 (\zeta) K_1 (\zeta) - \frac{\zeta^2}{2} [K_1^2(\zeta) - K_0^2 (\zeta)] \right], \label{eq:photonflux}
\end{align}
with $\alpha$ the electromagnetic coupling constant, $Z$ the atomic number, $\zeta = 2 x_\gamma M_p R_A$, $M_p$ the proton mass, and $R_A$ the nucleus radius. The transversely polarized cross section reads
\begin{align}
\frac{{d}\hat{\sigma}^T_{\gamma g \to q\bar{q}}}{{d}t} = -\frac{2\pi \alpha \alpha_s e_q^2}{s^2}.
\end{align}
Although both $\gamma g$ and $\gamma q$ contributes to the back-to-back dihadron production in photon-nucleus collisions, the dominant contribution arises from the connected $\gamma g$ channel \cite{Chen:2024qvx}. This is because that the photon flux drops exponentially at slight large $x_\gamma$, forcing $x_b$ to be very small as well. At small-$x_b$, the gluon distribution function is expected to be much larger than that of quark. Therefore, we expect that the experimental signal of dihadron transverse spin correlation in the UPC experiments is sizable.

Furthermore, in the future Electron Ion Collider (EIC) experiment, we can also perform the photon-nucleus scattering with a spacelike virtual photon. As presented in Ref.~\cite{Jezuita-Dabrowska:2002tjg}, the unphysical longitudinal photon also contributes, besides the physical transverse photon. Again, the transverse spin correlation only arises from the $\gamma^* g\to q\bar q$ channel, which is given by
\begin{align}
{\cal C}_{TT} = \frac{\int d{\cal P.S.} \sum_{{\cal S=L,T}} \sum_q G_{\gamma^*_{\cal S}} (x_{\rm Bj},Q^2) x_g f_{1,g}(x_g) \frac{1}{\pi} \frac{d\hat \sigma^T_{\gamma_{\cal S}^* g \to q\bar q}}{dt} H_{1T,q}^{\Lambda} (z_1) H_{1T,\bar q}^{\bar\Lambda} (z_2) }{\int d{\cal P.S.} \sum_{{\cal S=L,T}} \sum_{b,c,d} G_{\gamma_{\cal S}^*} (x_{\rm Bj},Q^2) x_b f_{1,b}(x_b) \frac{1}{\pi} \frac{d\hat \sigma_{\gamma_{\cal S}^* b\to cd}}{dt} D_{1,c}^{\Lambda} (z_1) D_{1,d}^{\bar\Lambda} (z_2)}.
\end{align}
Here $G_{\gamma_{\cal S}^*}(x_{\rm Bj},Q^2)$ is the photon flux in the DIS process \cite{Chen:2024qvx} with ${\cal S = L,T}$ representing the photon polarization. For the self-sufficient of this paper, we list the photon flux in Appendix \ref{sec:coeff}.  $x_{\rm Bj}$ is the Bjorken variable and $Q$ is the virtuality of the virtual photon. The transversely polarized partonic cross section can be expressed as
\begin{align}
&
\frac{d\hat\sigma^T_{\gamma_{\cal L}^* g \to q\bar q}}{dt} = \frac{2\pi \alpha \alpha_s e_q^2}{(s+Q^2)^2} \frac{4Q^2 s}{(s+Q^2)^2},
\\
&
\frac{d\hat\sigma^T_{\gamma_{\cal T}^* g \to q\bar q}}{dt} = - \frac{2\pi \alpha \alpha_s e_q^2}{(s+Q^2)^2} \left[ 1 - \frac{2Q^2 s}{(s+Q^2)^2} \right].
\end{align}
It is interesting to note that $d\sigma^T_{\gamma_{\cal L}^* g \to q\bar q}/d\sigma_{\gamma_{\cal L}^* g \to q\bar q} = +1$ indicating the unitary transverse spin correlation at arbitrary kinematics. The longitudinal photon leads to the positive correlation while the transverse photon leads to the negative correlation. Furthermore, when plotting $d\sigma^T_{\gamma_{\cal T}^* g \to q\bar q}/d\sigma_{\gamma_{\cal T}^* g \to q\bar q}$ as a function of $t/(s+Q^2)$, the ratio is $Q^2$ irrelevant. We obtain again the valley-shape correlation which is on a par with that of $gg\to q_i \bar q_i$ as a function of $t/s$. 

\section{Transverse spin correlation in the helicity amplitude approach}
\label{sec:helicity_amplitude}

The spin correlation can be better appreciated in the helicity amplitude approach. While it is quite obvious for the helicity correlation, it is a bit subtle for the transverse spin correlation. Nonetheless, it has been well presented in Refs.~\cite{Anselmino:2005sh, Anselmino:2011ch, DAlesio:2021dcx}. In this section, we summarize the essential idea on how to relate helicity amplitudes with the transversely polarized cross section. 

Let us consider the simple LO partonic scattering $a+b\to c+d$ with unpolarized beams. The helicity amplitude is thus denoted as ${\cal M}_{\lambda_a,\lambda_b} (\lambda_c, \lambda_d)$ with $\lambda_{a,b,c,d} = \pm 1$ denoting the helicity of the corresponding parton. The unpolarized cross section can simply be related to the helicity amplitudes \cite{Gastmans:1990xh} by 
\begin{align}
\frac{d\hat \sigma_{ab\to cd}}{dt} = \frac{1}{16\pi s^2} \frac{1}{4} \sum_{\lambda_a,\lambda_b,\lambda_c, \lambda_d} {\cal M}_{\lambda_a,\lambda_b}(\lambda_c, \lambda_d){\cal M}_{\lambda_a,\lambda_b}^* (\lambda_c, \lambda_d). 
\end{align}
The factor of $1/4$ averages the spin degree-of-freedom of the initial partons $a$ and $b$. The relation between the cross section and scattering amplitudes is taken from Ref.~\cite{Gastmans:1990xh}. When employing the above normalization, one needs to adopt the same convention for the scattering amplitude as that in Ref.~\cite{Gastmans:1990xh}. The transversely polarized cross section is expressed as
\begin{align}
\frac{d\hat \sigma^T_{ab\to cd}}{dt} = \frac{1}{16\pi s^2} \frac{1}{4} \sum_{\lambda_a,\lambda_b,\lambda_c,\lambda_d} {\cal M}_{\lambda_a,\lambda_b}(\lambda_c, \lambda_d){\cal M}_{\lambda_a,\lambda_b}^* (-\lambda_c, -\lambda_d). 
\label{eq:general-relation}
\end{align}
The transverse spin correlation in unpolarized collisions demands a sign flip for final-state-parton helicities in the amplitude and the conjugate amplitude, while maintaining the same helicities for initial state partons. If this requisite cannot be achieved, the transverse spin correlation disappears. Moreover, the positivity constrain $|d\sigma^{T}/d\sigma|<1$ is automatically satisfied, since the unpolarized cross section takes the form of $|{\cal M}_1|^2+|{\cal M}_2|^2$ and $\sigma^T$ is evaluated from $2{\rm Re}[{\cal M}_1 {\cal M}_2^*] \le 2 |{\cal M}_1||{\cal M}_2|$. This is thus fully consistent with the probability interpretation of polarized and unpolarized cross sections. Moreover, one can also derive $|{\cal C}_{TT}| \le \sqrt{1-|{\cal P}_L|^2}$ with $|{\cal P}_L|$ the magnitude of quark/antiquark helicity from the above relation. This positivity constrain reduces the transverse spin correlation at LEP experiment.

First, we use $e^+ e^- \to \gamma^* \to q\bar q$ to demonstrate the emergence of the transverse spin correlation. As presented in Ref.~\cite{Gastmans:1990xh}, the nonvanishing helicity amplitudes read
\begin{align}
&
{\cal M}_{+-} (+, -) =   e^2 e_q \sin\theta \sqrt{\frac{1+\cos\theta}{1-\cos\theta}},
&&
{\cal M}_{+-} (-, +) = - e^2 e_q \sin\theta \sqrt{\frac{1-\cos\theta}{1+\cos\theta}},
\label{eq:helicity-amp-1}
\\
&
{\cal M}_{-+} (+, -) = - e^2 e_q \sin\theta \sqrt{\frac{1-\cos\theta}{1+\cos\theta}},
&&
{\cal M}_{-+} (-, +) =   e^2 e_q \sin\theta \sqrt{\frac{1+\cos\theta}{1-\cos\theta}},
\label{eq:helicity-amp-2}
\end{align}
with $\theta$ the angle between momenta of $e^+$ and $q$. Substituting Eqs.~(\ref{eq:helicity-amp-1}-\ref{eq:helicity-amp-1}) into Eq.~(\ref{eq:general-relation}), the transversely polarized cross section is 
\begin{align}
\frac{d\hat\sigma^T_{e^+e^-\to\gamma^*\to q\bar q}}{dt} = \frac{N_c}{4} \frac{2{\rm Re}[{\cal M}_{+-} (+, -){\cal M}^*_{+-} (-, +) + {\cal M}_{-+} (+, -){\cal M}_{-+}^* (-, +)]}{16\pi s^2}
= -\frac{\pi \alpha^2 N_c e_q^2 \sin^2\theta}{Q^4}.
\end{align}
Utilizing $\sin^2 \theta = 4y(1-y)=2C(y)$ with $y=(1+\cos\theta)/2$ and $d\sigma/dy = Q^2 d\hat\sigma/dt$, we can recognize that it contributes to the electromagnetic term of $\omega_q^T$ in Eq.~(\ref{eq:omegaT}).

Second, we use $q_i q_j \to q_i q_j$ channel as an example to demonstrate the disappearance of transverse spin correlation. According to Ref.~~\cite{Gastmans:1990xh}, the scattering amplitudes in the helicity basis are given by
\begin{align}
&
{\cal M}_{++} (+,+) \neq 0, 
&&
{\cal M}_{++} (-,-) = 0, 
\\
&
{\cal M}_{+-} (+,-) \neq 0,
&&
{\cal M}_{+-} (-,+) = 0,
\\
&
{\cal M}_{-+} (-,+) \neq 0,
&&
{\cal M}_{-+} (+,-) = 0,
\\
&
{\cal M}_{--} (-,-) \neq 0,
&&
{\cal M}_{--} (+,+) = 0.
\end{align}
Since the sign flip cannot happen, the transverse spin correlation vanishes. Furthermore, another special case has already been discussed in Sec.~\ref{sec:ee}. When the coupling vertex takes the form of $\gamma^\mu(1\pm \gamma_5)$, the final state quark and antiquark are $100\%$ polarized along the helicity direction, leaving no phase space for a helicity flip. Therefore, the transverse spin correlation vanishes as required by the positivity constrain.

\section{Summary}

The quantitative research of spin-dependent fragmentation functions is way below satisfactory. Based on a series of studies \cite{Zhang:2023ugf, Li:2023qgj, Chen:2024qvx}, the spin correlation of back-to-back dihadron provides a complementary platform to understanding spin-dependent fragmentation functions in unpolarized high energy collisions. Therefore, the unpolarized colliders, currently available worldwide, have great potential to improve the status quo. 

In this paper, we focus on the transverse spin correlation of back-to-back hadrons in unpolarized  $e^+e^-$, $pp$, and $\gamma^{(*)} p$ collisions and investigate the opportunity of understanding the transverse spin transfer $H_{1T}$ in unpolarized high energy colliders. Unlike the case for the helicity correlation, the transverse spin correlation only resurfaces in the $q \bar q$ production channels and the identical-quark-scattering channel. Therefore, the magnitude of the transverse spin correlation in $e^+e^-$ and photon-nucleus collisions is expected to be sizable. In contrast, since $pp$ collisions contain substantial contributions from the unconnected diagrams, the magnitude of the transverse spin correlation is expected to be significantly reduced. To summarize, we have demonstrated the proof-of-concept that the unpolarized colliders can provide valuable information on the transverse spin transfer $H_{1T} (z)$. Future measurements of this observable can cast more light on the spin dependence of fragmentation functions.

\section*{Acknowledgments}
We thank Jian Zhou, Marco Zaccheddu, and Ya-Jin Zhou for fruitful discussion. This work is supported by Natural Science Foundation of China under grant No.~12405156, the Shandong Province Natural Science Foundation under grant No.~2023HWYQ-011 and No.~ZFJH202303, and the Taishan fellowship of Shandong Province for junior scientists.

\appendix

\section{Coefficient functions}
\label{sec:coeff}

In this appendix, we list all the coefficient functions used in this paper for the completeness of this paper. Some of those function can also be found in other references. 

For the electron positron annihilation process, those coefficient functions \cite{Wei:2013csa, Wei:2014pma, Chen:2016moq, Chen:2021zrr} are given by
\begin{align}
& \omega_q(y)=e_q^2 A(y)+\chi_{\text {int }}^q I_0^q(y)+\chi T_0^q(y), \\
& \omega_q^T (y) = - \left\{ e_q^2 + \chi_{\rm int}^q c_V^e c_V^q + \chi c_1^e [(c_V^q)^2-(c_A^q)^2] \right\} C(y), \label{eq:omegaT} \\
& T_0^q(y)=c_1^e c_1^q A(y)-c_3^e c_3^q B(y), \\
& I_0^q(y)=c_V^e c_V^q A(y)-c_A^e c_A^q B(y), \\
& A(y) = y^2 + (1-y)^2, \\
& B(y) = 1-2y, \\
& C(y) = 2y (1-y), \\
& \chi=\frac{Q^4} {[(Q^2-M_Z^2)^2+\Gamma_Z^2 M_Z^2] \sin ^4 2 \theta_W}, \\
& \chi_{\text {int}}^q=- \frac{2 e_q Q^2(Q^2-M_Z^2)}{[(Q^2-M_Z^2)^2+\Gamma_Z^2 M_Z^2] \sin ^2 2 \theta_W},
\end{align}
with $\theta_W$ the Weinberg angle, $c_1^e=(c_V^e)^2+(c_A^e)^2$, $c_3^e=2c_V^e c_A^e$, $c_1^q=(c_V^q)^2+(c_A^q)^2$, $c_3^q=2c_V^q c_A^q$, and $c_{V/A}$ the weak coupling constant \cite{Griffiths:2008zz}. For completeness, we list them in Table \ref{tab:numbers}.
{\renewcommand{\arraystretch}{1.5}
\begin{table}[h!]
\begin{tabularx}{0.45\textwidth} { 
  | >{\centering\arraybackslash}X 
 || >{\centering\arraybackslash}X 
  | >{\centering\arraybackslash}X | }
\hline
 & $c_V$ &  $c_A$  \\
\hhline{|=#=|=|}
$e$ & $-\frac{1}{2} + 2 \sin^2 \theta_W$ & $-\frac{1}{2}$ \\
\hline
$q=u, c, t$ & $\frac{1}{2} - \frac{4}{3}\sin^2\theta_W$ & $\frac{1}{2}$ \\
\hline
$q=d, s, b$ & $-\frac{1}{2} + \frac{2}{3} \sin^2\theta_W $ & $-\frac{1}{2}$ \\
\hline 
\end{tabularx}
\caption{Table for the coupling constants of weak interaction. Here, $\sin^2 \theta_W = 0.231$.}
\label{tab:numbers}
\end{table}

For the deep-inelastic scattering, the photon flux \cite{Caucal:2023nci} is given by
\begin{align}
&
G_{\gamma^*,{\cal L}} (x_{\rm Bj}, Q^2) = \frac{\alpha}{\pi Q^2 x_{\rm Bj}} (1 - y),
\\
&
G_{\gamma^*,{\cal T}} (x_{\rm Bj}, Q^2) = \frac{\alpha}{2\pi Q^2 x_{\rm Bj}} [1 + (1-y)^2],
\end{align}
with $y = Q^2/(x_{\rm Bj}s)$.

\section{The DGLAP evolution of $H_{1T} (z,\mu_f^2)$}
\label{sec:evolution}

The collinear $H_{1T} (z,\mu_f^2)$ fragmentation function follows the DGLAP evolution equation \cite{Altarelli:1977zs} with $\mu_f$ the factorization scale. Only quark/antiquark contributes, since the gluon fragmentation function disappears. The DGLAP evolution equation becomes diagonal, which is given by
\begin{align}
\frac{\partial H_{1T,q} (z,\mu_f^2)}{\partial \ln \mu_f^2} = 
\frac{\alpha_s (\mu_f^2)}{2\pi} \int_z^1 \frac{d\xi}{\xi} P_{qq}^T (\xi) H_{1T,q} (\frac{z}{\xi}, \mu_f^2).
\end{align}
Here, $P_{qq}^T$ is the transverse splitting function \cite{Artru:1989zv, Vogelsang:1997ak, Barone:2001sp} which, at the leading order, reads
\begin{align}
P_{qq}^T (\xi) = P_{qq} (\xi) - C_F (1-\xi) = C_F \frac{1+\xi^2}{(1-\xi)_+} + 2\delta (1-\xi)  - C_F (1-\xi).
\end{align}
The next-to-leading order result has also been derived in Ref.~\cite{Vogelsang:1997ak}.

Since the transverse splitting function, $P_{qq}^T$, is smaller than the unpolarized one, $P_{qq}$, it is obvious that the transverse polarization inherited by final state hadron becomes smaller after each splitting. This transverse spin loss due to the QCD evolution guarantees that the positivity constraint of spin dependent fragmentation function \cite{Soffer:1994ww, Vogelsang:1997ak} is automatically attained is as long as it is satisfied at the initial condition. We investigate the loss of transverse spin due to the QCD evolution in a toy model in the following.

\begin{figure}[h]
\centering
\includegraphics[width=0.3\textwidth]{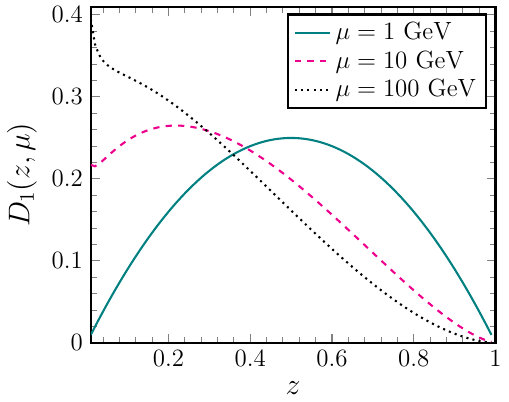}
\includegraphics[width=0.3\textwidth]{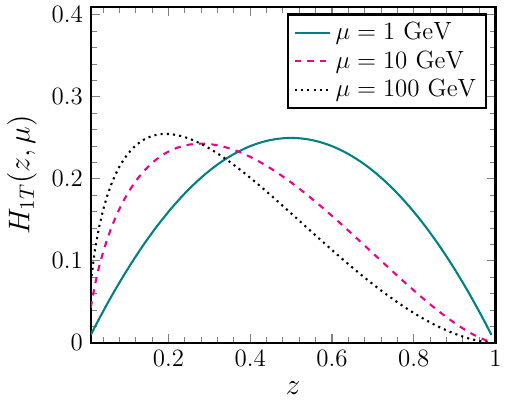}
\includegraphics[width=0.3\textwidth]{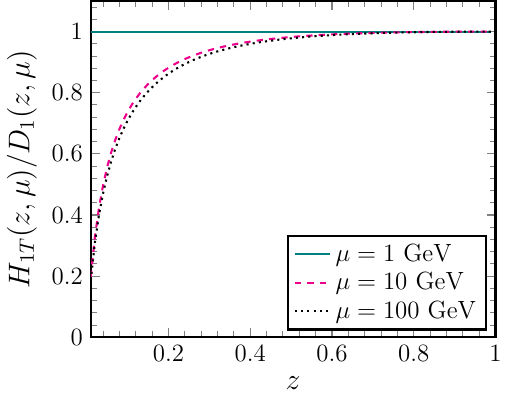}
\caption{Numerical results for the unpolarized and polarized fragmentation functions $D_{1}(z,\mu^2)$ and $H_{1T} (z,\mu^2)$ at $\mu = 1, 10$, and $100$ GeV.}
\label{fig:dglap-D}
\end{figure}

First, we parameterize the polarized and unpolarized fragmentation functions at the initial scale $\mu_0$ as $H_{1T,q} (z, \mu_0^2) = G_{1L} (z,\mu_0^2) = D_{1,q} (z, \mu_0^2) = z (1-z)$. The ratio $H_{1T}/D_1$ is unity at the initial scale, indicating that the transverse polarization of the quark $q$ is completely inherited by the final state hadron. Second, we only consider the diagonal terms in the DGLAP evolution for both functions, i.e., the unpolarized gluon fragmentation function is also neglected. The numerical results of $D_{1}$ and $H_{1T}$ at larger factorization scales are presented in Fig.~ \ref{fig:dglap-D}. Thanks to the DGLAP evolution, both $H_{1T}$ and $D_{1}$ decrease with increasing scale at large $z$, while they increase at small $z$. However, the increase of $H_{1T}$ at small $z$ is much slower than that of $D_{1}$ because of the transverse spin loss effect. 

The QCD evolution always suppresses the ratio $H_{1T}/D_1$. As shown in the right panel of Fig.~\ref{fig:dglap-D}, it rapidly decreases at small $z$ after a few steps of QCD evolution and eventually becomes stable at a large factorization scale. Furthermore, the transverse spin loss becomes negligible at the large $z$ region indicating the optimal kinematic region to measure this effect in high energy colliders. Notice that in this study with a toy model, we have also neglected the gluon contribution to the DGLAP evolution of $D_{1}$. Therefore, in reality, the loss of transverse polarization is even more severe. However, the qualitative feature remains unaltered. The polarization loss effect at small $z$ is more significant than that at large $z$ We leave a more sophisticated evaluation for a future study. 

Furthermore, this is a toy model calculation which is employed to demonstrate the transverse-spin-loss effect due to parton shower. In reality, the transverse spin transfer is required to satisfy a positivity constrain \cite{Soffer:1994ww, Vogelsang:1997ak} which reads
\begin{align}
|H_{1T,q} (z)| \le \frac{1}{2} \left[ D_{1,q} (z) + G_{1L, q} (z) \right]. 
\end{align}
According to the DSV parametrization \cite{deFlorian:1997zj} and the LEP data \cite{ALEPH:1996oew, OPAL:1997oem}, $G_{1L,q}^{\Lambda}(z) \sim z^\alpha D_{1,q}^{\Lambda} (z)$. Therefore, the ratio $H_{1T}/D_1$ can only reach unity at the threshold regime and it is expected to be smaller than $1/2$ at the $z\to 0$ limit. To summarize, the optimal kinematics of measuring the transverse spin correlation is the large $z$ region.

\section{Numerical Estimate of the transverse spin correlation in $pp$ collisions}
\label{sec:est}

In this appendix, we present a numerical estimate for the transverse spin correlation in $pp$ collisions. Notice that this is only an estimate based on a naive physical picture, since we barely know anything about the transverse spin transfer. In the naive picture, we assume $H_{1T,q} (z, \mu_0^2) = z D_{1,q} (z,\mu_0^2)$ for $q=u/d/s$ at the initial scale $\mu_0 = 1$ GeV, while sea partons do not contribute. 

\begin{figure}[htb]
\includegraphics[width=0.3\textwidth]{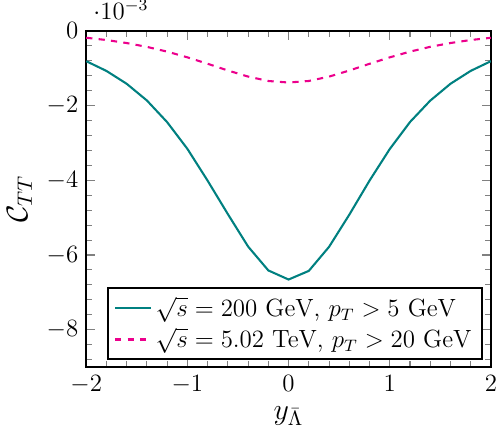}
\includegraphics[width=0.3\textwidth]{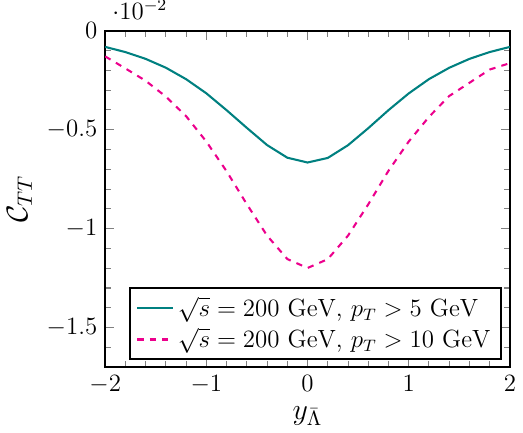}
\caption{Numerical estimates for the transverse spin correlation in $pp$ collisions based on a naive parameterization.}
\label{fig:pp-ctt}
\end{figure}

We thus present our numerical estimate for the transverse spin correlation of back-to-back $\Lambda/\bar\Lambda$ pair. The kinematics are chosen to be exactly the same with that in Ref.~\cite{Zhang:2023ugf}. The transverse momenta of $\Lambda$ and $\bar\Lambda$ have been integrated over and are required to be larger than certain value. We require the rapidity of $\Lambda$ to in the middle rapidity and show the transverse spin correlation as a function of $\bar\Lambda$ rapidity. The numerical estimates are shown in Fig.~\ref{fig:pp-ctt}. 

In the left panel, we show our numerical estimates for typical RHIC and LHC energies. Since the LHC energy is much higher that the RHIC energy, the dominate contribution comes from the small-$z$ region, where the spin transfer is assumed to be negligible. As a result, the dihadron transverse spin correlation is much smaller at the LHC energy. Furthermore, as a general feature, the transverse spin correlation becomes smaller in the forward/backward rapidity than that in the middle rapidity. This is mainly because that the $q+g$ channel dominates in the forward/backward rapidity, which does not exhibit transverse spin correlation.

In the right panel, we show the numerical estimates for RHIC energy with two different lower cutoffs. As expected, the one with a larger cutoff shows a stronger transverse spin correlation. Therefore, although the production rate for high $p_T$ dihadron is very small, the transverse spin correlation is very larger.

\end{document}